\documentclass[twocolumn,prl,aps]{revtex4}
\usepackage{graphics}
\begin{document}
\title{Apparent electron-phonon interaction in strongly correlated systems}
\author{O. R\"osch  and O. Gunnarsson}
\affiliation{ Max-Planck-Institut f\"ur Festk\"orperforschung,
D-70506 Stuttgart, Germany}

\begin{abstract}

We study the interaction of electrons with phonons in strongly correlated
solids, having high-$T_c$ cuprates in mind. Using sum-rules, we show that
the apparent strength of this interaction strongly depends on the property 
studied. If the solid has a small fraction (doping) $\delta$ of charge 
carriers, the influence of the interaction on the phonon self-energy is 
reduced by a factor $\delta$, while there is no corresponding reduction 
of the coupling seen in the electron self-energy. This supports the 
interpretation of recent photoemission experiments, assuming 
a strong coupling to phonons.  
\end{abstract}
\maketitle

There has been much interest in electron-phonon coupling in high-$T_c$ 
cuprates after photoemission spectroscopy (PES) studies of Lanzara
{\it et al.} \cite{Lanzara} showed a strong coupling to a mode at 
70 meV. This was interpreted as a coupling to a half-breathing phonon,
where O atoms in the CuO$_2$ plane move towards Cu atoms. This
interpretation is supported by the anomalous softening under 
doping \cite{Pintschovius1,Pintschovius2,Pintschovius3,anomol} and 
large width \cite{Pintschovius2} of this phonon. It is intriguing,
however, that a substantially larger apparent coupling strength $\lambda$
was estimated from a kink in the PES spectrum \cite{Lanzara} than 
what is suggested from the phonon width and softening \cite{coupling}. 
Similar considerations apply to the B$_{1g}$ phonon width \cite{Reznik} 
and the anomaly at 40 meV seen in PES \cite{Shen}. These estimates of 
$\lambda$ were based on theories which assume noninteracting electrons. 
Using sum-rules, we show that for a strongly correlated system the 
apparent $\lambda$ deduced from such theories should depend strongly on 
the property studied.  We consider a doped Mott (charge transfer) insulator, 
such as a cuprate, which has a fraction $\delta$ carriers (where $\delta$ 
typically is small).  The influence of the electron-phonon interaction on 
the phonon self-energy, determining its width and softening, is then 
reduced by a factor of the order of $\delta$ compared to a system without 
electron-electron interaction. 
For the electron self-energy, determining the photoemission spectrum,
there is no comparable reduction. This explains why the $\lambda$ deduced
from the phonon self-energy appears smaller than the one deduced from PES,
and it supports the scenario that phonons give a large contribution
to structures in PES.  For similar reasons, there should be no reduction 
$\sim \delta$ in the phonon induced carrier-carrier interaction. To address 
these issues, we present a method for determining $\lambda$ in exact 
diagonalization approaches.

The electron-electron interaction can strongly reduce the effects of
the electron-phonon interaction on phonons \cite{Jong}. The electron
density is rearranged in response to the excitation of a phonon,
and this rearrangement acts back on the phonon, contributing to
the width and energy shift of the phonon.  The response of the electrons
depends on electron hopping, which is hindered by interaction effects. In
particular, if the interaction is strong enough to lead to a Mott insulator
(for $\delta=0$), the electron-phonon contribution to the phonon width
vanishes. 

For cuprates, this can be studied in the $t$-$J$ model \cite{Zhang}, 
which has one site per Cu atom.  Each site is occupied by either a Cu 
$3d$-hole or a Zhang-Rice singlet, composed of a Cu $3d$-hole and an 
O $2p$-hole. The doping $\delta$ gives the fraction of singlets, which 
provide the charge carriers. The Hamiltonian is given by 
\begin{eqnarray}\label{eq:2.2}
H_{t\textrm{-}J}&=&
J\sum_{<i,j>}\left(
{\bf S}_i\cdot{\bf S}_j-\frac{n_in_j}{4}
\right) \nonumber \\
&-&t\sum_{<i,j>\sigma}(\tilde c_{i\sigma}^{\dagger}
\tilde c_{j\sigma}^{\phantom\dagger}+H.c.),
\end{eqnarray}
where $\tilde c_{i\sigma}^{\dagger}$ creates 
a $d$-hole on site $i$ if this site previously had no hole. A hole can 
hop with the hopping integral $t$ to sites occupied by singlets and vice 
versa. The spins of the $3d$-holes have a Heisenberg interaction with the 
interaction strength $J$. In the $t$-$J$ model with 
phonons \cite{Becker,Horsch1,Horsch2,Nagaosa1,Oliver,Nagaosa2}, the 
phonons couple mainly to the on-site energies and only weakly 
to the terms describing hopping and spin-spin interaction \cite{Oliver,Becker}. 
In the following we only include the coupling to the on-site term, 
\begin{equation}\label{eq:0}
H_{ep}={1\over \sqrt{N}}\sum_{i,{\bf q}} g_{\bf q}(n_i-1)(b_{\bf q}+
b_{-\bf q}^{\dagger})e^{i{\bf q}\cdot {\bf R}_i},
\end{equation}
where $N$ is the number of sites, $g_{\bf q}$ is a coupling constant, 
$n_i$ measures the $d$-hole occupation on the site at ${\bf R}_i$ and 
$b_{\bf q}$ annihilates a phonon with a wave vector ${\bf q}$. 

The phonons only couple to sites with Zhang-Rice singlets, i.e.,
sites with no $d$-holes. The phonon self-energy $\Pi({\bf q},\omega)$  
can then be expressed in terms of the charge-charge response function 
$\chi({\bf q}, \omega)$. 
\begin{equation}\label{eq:00}
\Pi({\bf q},\omega)={(g_{\bf q}^2/N)\chi({\bf q},\omega) \over
 1 + (g_{\bf q}^2/N)\chi({\bf q},\omega)D_0(
{\bf q},\omega)}, 
\end{equation}
where $D_0({\bf q},\omega)$ is the free phonon Green's function. 
Khaliullin and Horsch \cite{Horsch1} 
showed that there is a sum-rule 
\begin{equation}\label{eq:1}
{1\over \pi N}\sum_{{\bf q}\ne 0}\int_{-\infty}^{\infty} |{\rm Im}\chi(
{\bf q},\omega)|d\omega=2\delta(1-\delta)N,
\end{equation}
This result is a factor $2\delta(1-\delta)$
times the result for noninteracting electrons in a half-filled band.
Since $\chi({\bf q},\omega)$ becomes small for small $\delta$, the same 
is true for $\Pi({\bf q},\omega)$. In this limit the denominator in 
Eq.~(\ref{eq:00}) is not very important, and the sum-rule in Eq.~(\ref{eq:1})
also applies approximately to $\Pi({\bf q},\omega)/g_{\bf q}^2$. 

To understand the result in Eq.~(\ref{eq:1}), we notice that the system 
can respond to the perturbation of a phonon by transferring singlets to 
sites with Cu $3d$-holes. If there are few singlets, i.e., $\delta$ is small, 
the response of the system is weak and the phonon self-energy is small. 
Since typically $\delta\sim 0.1$, this drastically reduces the phonon 
softening and width. 

It is interesting to study PES and inverse PES (IPES) to see if many-body
effects also drastically change the effects of the electron-phonon interaction
in these cases. These spectra are described by the spectral function 
$A({\bf k},\omega)={\rm Im} G({\bf k},\omega-i0^{+})/\pi$, where 
${\bf k}$ is a momentum and $\omega$ a frequency.
$G({\bf k},z)$ is the electron Green's function
\begin{equation}\label{eq:2}
G({\bf k},z)= {a_{\bf k}\over z -\varepsilon_{\bf k}-
\Sigma({\bf k},z)},
\end{equation}
where $a_{\bf k}$ is a weight, $z$ is a complex number and $\Sigma({\bf k},z)$ 
is the electron self-energy. The $z$-independent part of $\Sigma$ is included
in the energy $\varepsilon_{\bf k}$, so that $\Sigma({\bf k},z)\sim b_{\bf k}/z$ 
for large $z$. To determine $b_{\bf k}$, we expand the Green's function in $1/z$
\begin{eqnarray}\label{eq:2a}
G({\bf k},z)&=&{a_{\bf k}\over z}(1+{\varepsilon_{\bf k}\over z}+{\varepsilon_{\bf k}^2 +
b_{\bf k} \over z^2} + ..)\\
&=&{a_{\bf k}\over z}(1+{\langle \omega \rangle_{\bf k} 
\over z}+ {\langle \omega^2 \rangle_{\bf k} \over z^2} + ..),\nonumber
\end{eqnarray}
where $\langle \omega^n\rangle_{\bf k}=\int 
\omega^n A({\bf k},\omega)d\omega/\int A({\bf k},\omega)d\omega$. 
The coefficient $b_{\bf k}$ can then be expressed in terms of $\langle \omega
\rangle_{\bf k}$ and $\langle \omega^2 \rangle_{\bf k}$. Using the analytical 
properties, we can relate $b_{\bf k}$ to a sum-rule over Im $\Sigma({\bf k},
\omega-i0^{+})$. Since there are no phonons in the ground-state in the model 
(\ref{eq:0}) for $\delta=0$, we can easily 
calculate expectation values of powers of $H_{ep}$, needed to obtain $\langle 
\omega^n\rangle_{\bf k}$. From this we deduce a sum-rule for the difference 
$\Sigma_{ep}$ between the self-energies with and without electron-phonon 
interaction.
\begin{equation}\label{eq:3}
{1\over \pi}\int_{-\infty}^{0}{\rm Im}\Sigma_{ep}({\bf k},
\omega-i0^{+})d\omega={1\over N}\sum_{\bf q} |g_{\bf q}|^2\equiv \bar g^2,
\end{equation}
which is valid for $\delta=0$. The integration only runs 
over energies corresponding to the PES spectrum, since the weight of the 
IPES spectrum is zero for $\delta=0$. The result (\ref{eq:3}) is equal to 
the lowest order result (in $g_{\bf q}^2$) for noninteracting electrons. 
While the effect of the electron-phonon coupling is strongly reduced by 
the small  doping for the phonon self-energy, there is no such reduction 
for the electron self-energy. Re $\Sigma_{ep}$ is related to Im
$\Sigma_{ep}$ via a Hilbert transform. 

The sum-rule in Eq.~(\ref{eq:3}) can be understood if we notice that a 
singlet with the wave vector ${\bf k}$, created in PES, can easily be scattered by 
phonons to other states ${\bf k+q}$, since only a fraction $(1-\delta)$ of these 
are occupied by other singlets. We therefore expect a strong electron-phonon 
interaction in PES also for a finite but small $\delta$.  As a result, one 
would then expect that the electron-phonon interaction appears to be a 
factor of $1/(c\delta)$ stronger in Re $\Sigma({\bf q},\omega)$ than for 
the phonon width 2 Im $\Pi({\bf q},\omega)$, where $c\sim 2-4$ depends 
on the assumptions about the $\omega$-dependencies of Im $\Sigma$ and 
Im $\Pi$. Although the arguments above show that the right hand side 
of the sum-rule in Eq.~(\ref{eq:3}) should not go to 0 for  
$\delta \to 0$, the result is, nevertheless, highly nontrivial.
The right hand side is independent of ${\bf k}$, $t$ and $J$. It is also
interesting that it remains proportional to $\bar g^2$ for large $\bar g$.

From these arguments it follows that the phonon induced attractive
carrier-carrier interaction should also be effective, since for small
$\delta$ the carriers (singlets) can scatter each other via phonons 
with few restrictions. This may be helpful for superconductivity. In 
particular, it should be possible to have a strong phonon induced 
carrier-carrier interaction, without the corresponding phonon going soft.

To illustrate these points, we study the $t$-$J$ model with phonons, 
using exact diagonalization. We consider a finite cluster with $4\times 3$ 
sites, and include the entire breathing phonon branch. The Hilbert space
is limited by only allowing for states which have a maximum of $K$ phonons, 
where typically $K=7$.  Ideally, we would calculate the PES spectrum as  
a function of ${\bf k}$. We would then expect to see a kink in the 
dispersion at energies of the order of $\omega_{ph}$ away from the 
Fermi energy, where $\omega_{ph}$ is a typical phonon energy. For the 
small clusters which can be studied in exact diagonalization, this 
approach is not possible, due to the few ${\bf k}$-points available. 

Instead we focus directly on the self-energy. The spectral function 
$A({\bf k},\omega)$ is calculated using exact diagonalization. The 
Green's function is then obtained from a Kramers-Kronig transformation. 
Finally, $\Sigma({\bf k},\omega)$ is obtained by inverting 
Eq.~(\ref{eq:2}). In a similar way we determine $\Pi({\bf q},\omega)$ 
from the phonon spectral function. The $\omega$-dependence of Re 
$\Sigma({\bf q},\omega)$ then gives information about the kink in 
PES and Im $\Pi({\bf q},\omega_{ph})$ gives the phonon width.

\begin{figure}
\centerline{
{\rotatebox{0}{\resizebox{8.5cm}{!}{\includegraphics {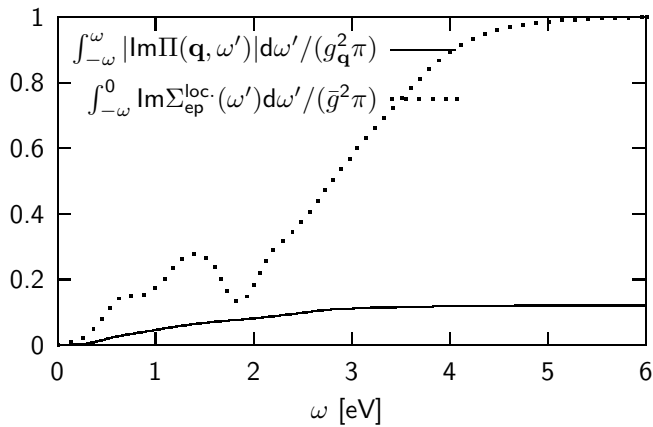}}}}}
\caption{\label{fig:1} 
Frequency integrals over the imaginary parts of the phonon self-energy
for ${\bf q}=\pi/(2a)(1,0)$ and the electron-phonon contribution to the 
${\bf k}$-averaged electron self-energy. The results are divided by the coupling 
constants and they were obtained for a $4\times 3$ cluster with 
periodic boundary conditions. $\Pi$ was calculated for $\delta=1/12$ and
$\Sigma_{ep}$ for $\delta=0$. The self-energies were given a 0.2 eV (FWHM)
Lorentzian broadening. The figure illustrates how these quantities converge
to approximately $2\delta\approx 0.17$ [Eq.~(\ref{eq:1})] and
unity [Eq.~(\ref{eq:3})] for the phonon and electron self-energy,
respectively.}
\end{figure}

Due to the small size of the cluster there are few many-electron 
states in the phonon energy range, which makes it hard to extract 
phonon widths, even in the approach above. The main points of the paper can, 
however, also be illustrated by using a larger phonon energy. We have 
therefore increased the bare phonon energy to $\omega_{ph}^0=0.5$ eV.
This requires a corresponding increase in the coupling constant. We have 
chosen a multiplying factor of 3.4, which leads to an apparent 
coupling strength to the phonons of the order seen experimentally \cite{coupling1}. 
We have furthermore used the parameters $t=0.47$ eV and $J/t=0.3$.

Figure~\ref{fig:1} shows the frequency 
integrals of the imaginary parts of the electron and phonon self-energies 
as a function of the upper limit $\omega$. The limit $\omega\to \infty$ 
corresponds to the sum-rules (\ref{eq:1}) and (\ref{eq:3}).  To obtain  
dimensionless quantities, the coupling constants have been divided out. 
The figure illustrates the large ratio, $\sim 1/(2\delta)$, of the 
sum-rules. The sum-rule in Eq.~(\ref{eq:1}) applies to an average 
over ${\bf q}\ne 0$ of $\chi$, but we found that it is also rather accurate 
for $\Pi/g_{\bf q}^2$ for an individual value of ${\bf q}$, as is illustrated
in Fig.~\ref{fig:1}.
The reason is that the denominator in Eq.~(\ref{eq:00}) is not very important
for the coupling strengths used here and that the sum-rule for an individual
${\bf q}$ is not very different from the average over ${\bf q}$.

\begin{figure}
\centerline{
{\rotatebox{0}{\resizebox{8.5cm}{!}{\includegraphics {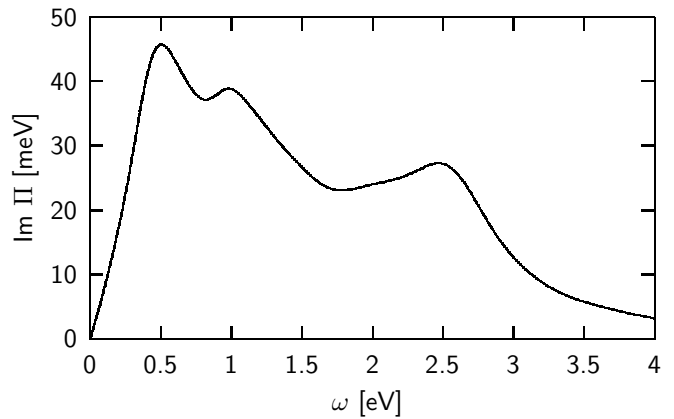}}}}}
\caption[]{\label{fig:2}
Im $\Pi({\bf q},\omega)$ of a $4\times 3$ cluster
for ${\bf q}=\pi/(2a)(1,0)$ and $\delta=1/12$. The self-energy has been
given a Lorentzian broadening of 0.4 eV.}
\end{figure}

The coupling strength $\lambda$ is often determined from the 
phonon width $\gamma=2\ {\rm Im}\ \Pi({\bf q},\omega_{ph})$ \cite{Allen} 
or its softening, $\Delta \omega_{ph}={\rm Re}\ \Pi({\bf q},\omega_{ph})$, using
\begin{equation}\label{eq:000}
\lambda^{\Pi} = {\gamma\over 2\pi \omega_{ph}^2N(0)}= -\alpha{\Delta 
\omega_{ph}\over \omega_{ph}^0},
\end{equation}
appropriate for noninteracting electrons. Here $N(0)$ is the electron 
density of states per spin and $\alpha \sim 1$ depends on the precise 
$\omega$ dependence of Im $\Pi({\bf q},\omega)$. Figure~\ref{fig:2} shows 
the broadened Im $\Pi({\bf q},\omega)$. The broadening (0.4 eV FWHM) was 
chosen in such a way that the fine structures due to the finite cluster
size were removed and so that the expected behavior Im $\Pi({\bf q},\omega)
\sim \omega$ for small $\omega$ was obtained. The phonon is softened to 
$\omega_{ph}=0.4$ eV, due to the electron-phonon interaction. For this 
frequency we obtain the FWHM of the phonon as $\gamma=0.08$ eV. This result 
depends on the broadening of Im $\Pi({\bf q},\omega)$ and results differing
by $\pm  30 \%$ could be obtained for other reasonable broadenings. 
Based on the width of Im $\Pi({\bf q},\omega)$, we estimate $N(0)\sim 0.5$
states per eV and spin and $\lambda^{\Pi}=0.2$. From the phonon 
softening $\Delta \omega_{ph}/ \omega_{ph}^0=0.2$, we obtain a similar 
result for $\lambda^{\Pi}$.  

We emphasize the difference in this approach, using a broadened $\Pi$,
from an approach where the phonon spectral function is broadened
until a smooth spectrum is obtained. The latter approach leads to 
an additional width of the peaks. For the present system, such a large 
broadening would be required that the approach would not be meaningful. 
The broadening of Im $\Pi({\bf q},\omega)$, on the other hand, does not 
generally add to the width of the phonon spectral function, since it 
essentially only distributes the contributions to Im $\Pi({\bf q},\omega)$ 
more uniformly in $\omega$. 

We next consider the coupling strength seen in the electron self-energy
determining PES. We can determine a $\lambda^{\Sigma}$ as 
\begin{equation}\label{eq:4}
\lambda^{\Sigma}=- \left. {d{\rm Re} \Sigma_{ep}(\omega) \over d\omega}
\right|_{\omega=0}= {1\over \pi} \int d\omega 
{{\rm Im}\Sigma_{ep}(\omega)\over \omega^2}.
\end{equation}
This leads to $\lambda^{\Sigma}=0.6$, which is about a factor of $1/(c\delta)$ 
larger than $\lambda^{\Pi}$ with $c=4$. As can be seen from Fig.~\ref{fig:1}
Im $\Sigma$ is small for $|\omega|<\omega_{ph}=0.4$ eV. For this 
frequency range we find that Re $\Sigma({\bf k},\omega)\sim 
-\lambda^{\Sigma}\omega$.

To summarize, we have found that in strongly correlated systems
the apparent strength $\lambda$ of the electron-phonon coupling 
crucially depends on the property of interest. For the $t$-$J$ 
model of a high-$T_c$ cuprate with a fraction $\delta$ carriers, 
sum-rules for the imaginary parts of the electron and phonon 
self-energies show a reduction by a factor $\delta$ for the phonon but 
not the electron case. This suggests that the apparent  $\lambda$ deduced from 
phonon widths and softenings is reduced by such a factor, while there is no 
reduction in the electron self-energy. This provides support for phonons being 
essential for kinks seen in photoemission. Similar arguments suggest
that the phonon induced 
interaction between the carriers is not reduced by a factor of $\delta$, 
which may be of importance for superconductivity. 
 
We thank O. K. Andersen, P. Horsch, B. Keimer, N. Nagaosa, D. J. Scalapino,
Z.-X. Shen and R. Zeyher for many useful discussions, and Z.-X. Shen for his
hospitality during a visit to Stanford.


\begin{thebibliography}{99}


\bibitem{Lanzara}A. Lanzara, P. V. Bogdanov, X. J. Zhou, S. A. Kellar, 
D. L. Feng, E. D. Lu, T. Yoshida, H. Eisaki, A. Fujimori, K. Kishio, 
J.-I. Shimoyama, T. Noda, S. Uchida, Z. Hussain, and Z.-X. Shen,
Nature {\bf 412}, 510 (2001).

\bibitem{Pintschovius1}S. L. Chaplot, W. Reichardt, L. Pintschovius,
and N. Pyka, Phys. Rev. B {\bf 52}, 7230 (1995); L. Pintschovius, 
N. Pyka, W. Reichardt, A. Y. Rumiantsev, N. L. Mitrofanov, A.S. Ivanov,
G. Collin, and P. Bourges, Physica (Amsterdam) {\bf C185-C189}, 156 (1991).

\bibitem{Pintschovius2}L. Pintschovius and M. Braden, Phys. Rev. B 
{\bf 60}, R15039 (1999).

\bibitem{Pintschovius3}L. Pintschovius and W. Reichardt, in {\it Neutron
Scattering in Layered Copper-Oxide Superconductors}, A. Furrer, Ed.,
Physics and Chemistry of Materials with Low Dimensional Structures,
Vol. 20 (Kluwer Academic, Dordrecht, 1998), p. 165.

\bibitem{anomol}R. J. McQueeney, Y. Petrov, T. Egami, M. Yethiraj,
G. Shirane, and Y. Endoh, Phys. Rev. Lett. {\bf 82}, 628 (1999).

\bibitem{coupling}Using Eq. (\ref{eq:000}), a calculated density of 
states \cite{Mattheiss} and measured values of the (half-)breathing phonon 
width \cite{Pintschovius2} and softening one obtains the apparent coupling 
constant for the phonon self-energy $\lambda^{\Pi} \sim 0.1-0.3$ for $\delta
=0.15$. While the apparent coupling for the electron self-energy $\lambda
^{\Sigma} \sim 1$ was originally estimated from PES, later work assigned 
a substantial part of this coupling to lower-lying phonons \cite{spectral}, 
bringing $\lambda^{\Sigma}$ and $\lambda^{\Pi}$ for the (half-)breathing 
phonon into better agreement. The observed kink in the PES spectrum can, 
however, substantially underestimate $\lambda^{\Sigma}$ in Eq. (\ref{eq:4}). 
Let the electron self-energy $\Sigma(\omega)=-(\lambda^{el}+\lambda^{\Sigma})
\omega$ for small $|\omega|$ and $\Sigma(\omega)=-\lambda^{el}\omega$
for large $|\omega|$, where $\lambda^{\Sigma}$ is due to the phonon of 
interest and $\lambda^{el}$ is due to other effects. Then the observed
change of slope at the kink is determined by $1+\lambda^{\Sigma}
/(1+\lambda^{el})< 1+\lambda^{\Sigma}$, where $(1+\lambda^{\Sigma}+
\lambda^{el})=Z^{-1}$, the inverse quasiparticle weight, can be a large
number. This suggests that $\lambda^{\Sigma}$ is substantially larger 
than $\lambda^{\Pi}$.

\bibitem{Reznik}D. Reznik, B. Keimer, F. Dogan, and I. A. Aksay,
Phys. Rev. Lett. {\bf 75}, 2396 (1995).

\bibitem{Shen}T. Cuk, F. Baumberger, D. H. Lu, N. Ingle, X. J. Zhou, 
H. Eisaki, N. Kaneko, Z. Hussain, T. P. Devereaux, N. Nagaosa, and
Z.-X. Shen, cond-mat/0403521.

\bibitem{Jong}J. E. Han and O. Gunnarsson, Phys. Rev. B {\bf 61}, 8628
(2000).

\bibitem{Zhang}F. C. Zhang and T. M. Rice, Phys. Rev. B {\bf 37}, 3759 (1988).

\bibitem{Becker}K. J. von Szczepanski and K.W. Becker, Z. Phys. B
{\bf 89}, 327 (1992).

\bibitem{Horsch1}G. Khaliullin and P. Horsch, Phys. Rev. B {\bf 54},
R9600 (1996).

\bibitem{Horsch2}P. Horsch, G. Khaliullin, and V. Oudovenko, Physica C 
{\bf 341}, 117 (2000).

\bibitem{Nagaosa1}Z.-X. Shen, A. Lanzara, S. Ishihara, and N. Nagaosa, 
Philos. Mag. B {\bf 82}, 1349 (2002).

\bibitem{Oliver}O. R\"osch and O. Gunnarsson, Phys. Rev. Lett. {\bf 92}, 
146403 (2004).

\bibitem{Nagaosa2} S. Ishihara and N. Nagaosa, Phys. Rev. B {\bf 69}, 
144520 (2004).

\bibitem{coupling1}We expect that keeping $g_{\bf q}^2/\omega_{ph}$
fixed when changing $\omega_{ph}$ would lead to an unchanged $\lambda$. 
We have multiplied the $g_{\bf q}$ rescaled in this way by an additional 
factor of $\sqrt{2}$. Deducing $\lambda$ from the phonon width then gives 
a value comparable to what is obtained from experiment. 

\bibitem{Allen}P. B. Allen, Phys. Rev. B {\bf 6}, 2577 (1972);
Solid State Commun.  {\bf 14}, 937 (1974).

\bibitem{Mattheiss}L. F. Mattheiss, Phys. Rev. Lett. {\bf 58}, 1028 (1987).

\bibitem{spectral}X. J. Zhou, J. Shi, T. Yoshida, T. Cuk, W. L. Yang,
V. Brouet, J. Nakamura, N. Mannella, S. Komiya, Y. Ando, F. Zhou, W. X. Ti,
J. W. Xiong, Z. X. Zhao, T. Sasagawa, T. Kakeshita, H. Esaki, S. Uchida,
A. Fujimori, Z. Zhang, E. W. Plummer, R. B. Laughlin, Z. Hussain,
and Z.-X. Shen,  cond-mat/0405130.


\end{thebibliography}
\end{document}